\documentclass[12pt]{article}                            
\usepackage{psfrag}
\usepackage{graphicx}                                                                                

\title{Performance of Wang-Landau algorithm in continuous spin models and a case study : modified XY-model\\[8mm]
} 
\author{Suman Sinha
\footnote{E-mail: ssinha@research.jdvu.ac.in}
 ~and 
Soumen Kumar Roy
\footnote{Corresponding author. E-mail: skroy@phys.jdvu.ac.in,
Tel: +91 9331910161; fax: +91 33 24146584}\\ 
Department of Physics,\\ 
Jadavpur University, Kolkata - 700032, INDIA}
\date{  }

\begin{document}

\maketitle
\begin{abstract}
Performance of Wang-Landau (W-L) algorithm in two continuous spin models is tested by determining the
fluctuations in energy histogram. Finite size scaling is performed on a modified XY-model using different
W-L sampling schemes. Difficulties faced in simulating relatively large continuous systems using W-L 
algorithm are discussed.
\end{abstract}
{\it PACS:} 05.10.Ln, 05.70.Fh, 64.60.Cn\\
{\it keywords:} Monte Carlo, Wang-Landau, Finite Size Scaling
\\
\\

The Wang-Landau (W-L) algorithm for Monte Carlo (MC) simulation, introduced in 2001
 \cite{wl} has since been applied to a wide range of problems in statistical physics
 \cite{va1,va2,va3,va4,jaya}. In most of these investigations the authors have applied 
the W-L algorithm to systems with discrete energy levels which include the Ising and the Potts
model. Relatively fewer papers have so far appeared on continuous models
\cite{cont,local,cont1} and in such systems one uses a discretization scheme to divide the energy range
of interest into a number of bins which label the macrostates of the system.

In the W-L algorithm, one directly determines the density of states $\Omega(E_{i})$
(in practice, its logarithm $g(E_{i})$) of the system with i=1,2,..n being the bin index.
 These macrostates 
are sampled with a probability which is inversely proportional to the current value
of the density of states. If a trial move passes the probability test, $g(E_{j})$ for the new
energy $E_{j}$ is modified as $g(E_{j}) \rightarrow  g(E_{j}) +\ln f$, where the modification factor
$f$ is $ \geq 1$ in the beginning of the simulation. In case a trial move fails, the density of states 
corresponding to the old value $E_{i}$ of the energy is modified. A histogram record $H(E_{i})$ of
all states visited is maintained throughout the simulation. When the $g(E_{i})$ corresponding to a 
certain macrostate is modified by adding $ \ln f$ to it, the corresponding $H(E_{i})$ is modified as
$H(E_{i}) \rightarrow H(E_{i}) +1$. In the beginning of the simulation the $g(E_{i})$'s for all macrostates
are initialized to zero. In the original version of W-L algorithm, an iteration is considered to be 
complete when the histogram satisfies a "flatness" criterion. This means that $H(E_{i})$ for all values of $i$
, has attained $90\%$ (or some other preset value) of the average of all $H(E_{i})$. In the following iteration
$f$ is reduced, the $H(E_{i})$'s are reset to zero, and the process is continued till $\ln f$ is as small as
$10^{-8}$ or $10^{-9}$. Since the history of the entire sampling process determines the density of states, the 
W-L algorithm is non-Markovian besides being multicanonical in nature.

In course of the random walk in a W-L simulation, the fluctuation of the energy histogram, for a given 
modification factor $f$, initially grows with the number of Monte Carlo sweeps and then saturates to a 
certain value. Because of the nature of the W-L algorithm, which has been described above, the value of the
histogram fluctuation determines the error which is generated in the resulting density of states. Zhou
and Bhatt \cite{zhou} carried out a mathematical analysis of the W-L algorithm. They proved the convergence 
of the iterative procedure and have shown that the error in the density of states, for a given $f$, is 
of the order of $\sqrt(\ln f)$. This finding has been tested by Lee $et. al$ \cite{lol} who performed
extensive numerical test in two discrete models. In Monte Carlo simulation
on two two-dimensional Ising models, namely, the ferromagnetic Ising model (FMIM) and the fully frustrated Ising
model (FFIM), Lee $et. al$ have shown that the fluctuation in the histogram increases during an initial
accumulation stage and then saturates to a value which is inversely proportional to $\sqrt(\ln f)$ and they
were of the view that this feature is generic to the W-L algorithm. 
As is
shown in references \cite{zhou} and \cite{lol} and also in the later part of this Letter, the resulting
error in the density of states is then of the order of $\sqrt(\ln f)$, which is in agreement with the prediction
of Zhou and Bhatt.

Methods which are alternatives to the requirement of the histogram flatness, for deciding where to stop
an iteration for a given modification factor, have been proposed in the work described in references 
\cite{zhou} and \cite{lol}. Zhou and Bhatt \cite{zhou} were of the opinion that an iteration may be
stopped when the minimum number of visits to each macrostate is $1/\sqrt(\ln f)$. On the other hand,
Lee $et. al$ \cite{lol} proposed that an iteration can be stopped when the number of Monte Carlo sweeps,
 for a given value of $f$, is such that the saturation of the histogram fluctuation has been reached, since
continuing the simulation for this particular modification factor is unlikely to reduce the error in the
density of states any further.
Subsequently, there have been a number of proposed improvements and studies of the efficiency
and convergence of this algorithm \cite{im1,im2,im3}.

	In this Letter we first present the results of our investigation on the growth of histogram
fluctuations in two continuous models. One of our aims is to check if the conjecture of Lee $et.al.$, that the 
nature of the dependence of the maximum of the histogram fluctuation on the modification factor $f$ is model
independent, can be extended to lattice spin models with continuous energy spectrum.
For this purpose, we have chosen a two dimensional spin system, where the spins,
 confined to a square lattice and free to rotate in a plane, 
say the xy- plane, (having no z-component) interact with the nearest neighbours via a
potential,
\begin{equation}
V(\theta)=2\left[1-\left(cos^{2}{\frac{\theta}{2}}\right)^{p^2}\right]
\end{equation}
where $\theta$ is the angle between the interacting spins and $p^{2}$ is a parameter to be chosen. 
 This model, now known as the modified
XY- model, was first introduced by Domany $et.al$ in 1984 \cite{ds}. For $p^{2}=1$, the model is simply the conventional
two-dimensional XY- model, which is known to exhibit a quasi-long-range-order-disorder 
 phase transition mediated by the unbinding of
topological defects \cite{him}. For large values of $p^2$, the potential of eqn.(1) has a sharp well structure
for small values of $\theta$ and the model exhibits a rather strong first order phase transition. Thus, the 
model with interaction determined by eqn.(1), for values of $p^2=1$ and $50$, effectively behaves like two
different models, having characteristically different phase transitions, although the symmetry of the 
Hamiltonian and the lattice and the spin dimensionalities remain the same.
A similar change in the nature of the phase transition has been observed in the two dimensional Lebwohl-Lasher (LL)
model and a modified version of it \cite{apskr}. In this model, with spin dimensionality $n=3$ and lattice dimensionality $d=2$, the
nearest neighbour spins interact via a potential $-P_2(cos \theta_{ij})$ and in the modified (LL) model
this is  $-P_4(cos \theta_{ij})$ where, $P_2$ and $P_4$ are the second and the fourth Legendre polynomials respectively.
Both models have the $O(3)$ as well as the local $Z_2$ symmetry. But as the nature of the interacting potential is 
modified, the transition changes from continuous to a sharply first order one. It has also been observed that for the $n=3$
and $d=3$ LL model, if one adds a $P_4$ interaction to the usual $P_2$ term,  
 the nature of the phase transition changes from a weakly first order one to one where the
discontinuities, which characterize the first order transition, grow sharply \cite{pel,mr1,mr2,mr3,mr4,mr5}.

 The $p^{2}=50$ model
 has very large fluctuations in energy, which is manifested in a huge peak in the specific heat. This is accompanied
 by very long relaxation times that make it difficult to obtain good statistics in the simulation using the
conventional canonical sampling as is done in the Metropolis method \cite{metro}.
 In their original work, Domany $et.al$ \cite {ds} had to carry out the simulation of this model in the roughening 
representation \cite{ro1} and employ long runs.

Using different approaches mentioned below, we have also tested the performance of the W-L algorithm in
simulating the $p^2=50$ model. In the first method, which we call 'a', a pre-assigned number of MC Sweeps,
determined from the saturation of the histogram fluctuation, is used to stop an iteration with a given
modification factor. In the other two methods, 'b' and 'c', a minimum number of visits to each macrostate
(bins) namely $1/\sqrt(\ln f)$ and $1/(\ln f)$ respectively have been used. So, in our work three different criteria,
instead of testing the flatness of the histogram, have been used for stopping a particular iteration.
Besides the methods 'a' and 'b' suggested in references \cite {lol} and \cite {zhou} respectively, we have
chosen method 'c' because the proposition of $1/\sqrt(\ln f)$
visits to each macrostate does not directly follow from the fact that the error in the density of states
is of the order of $\sqrt(\ln f)$. We have exercised this option, although it leads to a huge increase
in the amount of computation, to see in particular if it leads to any improvement in the results 
of simulation. This is necessary particularly in view of the fact that, we have worked in a continuous model and
the number of microstates which correspond to each bin is enormously large.

Before presenting our results we explain in detail the notations and symbols relevant to the problem.
We represent by $\beta_k$, the saturation value of the energy histogram fluctuation in the $k^{th}$ iteration.
 Let $ f_{k}$ be the modification factor for the 
$k^{th}$ iteration. One usually starts with a modification factor $ f=f_{1} \geq 1$ and uses a sequence of decreasing
$ f_{k}$'s (k=1,2,3,....) defined in some manner. 
 One MC sweep is taken to be completed when the number of attempted single particle
moves equals the number of particles in the system.  
The error in the density of states after the $n^{th}$ iteration has been performed, 
is directly related to $ \beta_{i}$ for $ i > n$, the saturation values of the fluctuations.  

	In the W-L algorithm the logarithm of the density of states after n iterations is given by
\begin{equation}
 g_{n}(E_{i})= \sum_{k=1}^n H_{k}(E_{i}) ~\ln(f_{k})
\end{equation}
where, $H_k(E_i)$ is the accumulated histogram count for the $i^{th}$ energy bin during the $k^{th}$ iteration.
 In order to get an idea of
the fluctuations in the histogram and its growth with the number of MC sweeps we subtract the minimum
of the histogram count $h_{k}^{j}$ which occurs in the histogram after the $j^{th}$ MC sweeps has been completed during
the $k^{th}$ iteration, $i.e$ we consider the quantity

\begin{equation}
\widetilde H_k^j (E_i) = H_k^j(E_i) - h_k^j
\end{equation}
(It may be noted that $h_k^j$ does not refer to any particular bin and may occur randomly in any of the bins).

	The quantity $\widetilde H_{k}^{j}(E_i)$ is now summed over all bins to give $\Delta H_k^j$.
\begin{equation}
\Delta H_k^j = \sum_i \widetilde H_k^j(E_i)
\end{equation}
$\Delta H_k^j$ is thus a measure of the fluctuations which occurs in the $j^{th}$ MC sweeps during $k^{th}$ iteration and is a
sort of average over all macrostates or bins. 
$\Delta H_k^j$ fluctuates with $j$ because of statistical errors and its mean value taken over $j$ is nothing but 
$\beta_k$.
The error in the logarithm 
of the density of states, summed over all energy levels or bins, after the completion of n iterations is therefore given by
\begin{equation}
\eta_n = \sum_{k=n+1}^{\infty} \beta_k ~ \ln(f_k)
\end{equation}
Since the predicted value of the error $\eta_n$ is of the order of $\sqrt(\ln f_n)$ \cite {zhou}, one expects
that the histogram saturation value $\beta_n$, for the $n^{th}$ iteration, should be proportional to
$1/\sqrt(\ln f_n)$.

	In the present work we have determined for the two lattice models we have defined, the dependence
of the quantity $\Delta H_k^j$, given by eqn(4), on j, the number of MC sweeps for a given iteration denoted by k.
Besides this, by using lattices of linear dimension $L=8$ to $80$ for the $p^2=50$ model we have performed finite
size scaling (FSS) using methods 'b' and 'c' so that the relative performance of the two methods can be judged.

For both the models and for the purpose of testing the fluctuation in histograms
we have worked on two system sizes, namely $8\times8$ and $16\times16$, and nearest neighbour interactions along with
periodic boundary condition were always used. The starting value of the modification factor  $\ln f_1$ was taken to be
0.1 and the sequence $\ln f_{n+1}=\ln f_n/(10)^{1/4}$ was chosen and for the purpose of determination of fluctuations
the minimum $\ln f$ used was $10^{-5}$. Clearly, the chosen sequence of $f$ is to ensure that it gets reduced 
by a factor of 10 after four iterations.  We have determined the quantity $\Delta H_k^j$ defined by eqn(4) at intervals
of $10^3$ MC sweeps and the maximum number of sweeps chosen for a given value of $f$ is such that the saturation of the
histogram is clearly evident.  
The energy range accessible to the systems we have investigated is 0 to 4 per spin. To reduce the computer time we 
have taken the system energy range to be 3 to 253 for the $8\times8$ lattice and 10 to 1004 for the $16\times16$ one. The width of
the energy bins was taken to be 0.2 in all cases.

In fig. 1, we have plotted the fluctuations in the histogram $\Delta H_k^j$ against the number of MC sweeps, j for four
values of the refining factor $f$. The plots shown are for the $16\times16$ lattice for $p^2=50$ and
for $\ln f$ equal to $10^{-2}$, $10^{-3}$, $10^{-4}$, $10^{-5}$. We did not go to values of $\ln f$ less than 
$10^{-5}$ as it takes a very large CPU time. Averages were taken over fifty independent simulations to improve the 
statistics. Similar plots were also taken for the $p^2=1$ model. 
It is evident from figure 1 that $\Delta H_k^j$ increases initially and then saturates and as $f$ gets smaller, the
saturation value as well as the number of MC sweeps necessary to reach the saturation increases. The standard
errors calculated from the fifty independent simulations are also shown.
In fig. 2 we have plotted the logarithm of the saturation value, 
$\beta_k$ $i.e$ $\ln(~\beta_k)$ vs $\ln(ln f)$ for the $p^2=1$ model for $8\times8$ and $16\times16$ system sizes and those for the $p^2=50$ 
model are depicted in fig. 3. The error bars are not shown since they are of size smaller than the dimension
of the symbols used for plotting.
From  these figures it is clear that
\begin{equation}
\beta_k \propto (\ln f)^{\alpha}
\end{equation}
where the index $\alpha=-0.508 \pm 0.002$ and $-0.507 \pm 0.002$ for $8\times 8$ and $16\times 16$ respectively
 for $p^2=1$ and $\alpha=-0.500 \pm 0.003$ and $-0.504 \pm 0.003$ for $8\times 8$ and $16\times 16$ respectively
for $p^2=50$. This is in agreement with the prediction of Zhou and Bhatt \cite{zhou} and the findings of
Lee $et.al$ \cite{lol} for the two discrete Ising models.

We denote by $\lambda$ the number of MC sweeps necessary to reach the saturation value of the histogram fluctuation.
Our results show that $\ln \lambda$ is a linear function of $\ln (\ln f)$, like $\ln \lambda=a+b\ln(\ln f)$
\begin{equation}
i.e., ~~~~~~\lambda=exp(a)(\ln f)^b
\end{equation}
where a, b are parameters. We have plotted $\ln \lambda$ against $\ln (\ln f)$ in figure 4 for the $p^2=50$ for
two lattice sizes. The fits
yield $a=8.78 \pm 0.18$ and $b=-0.526 \pm 0.02$ for the $8\times 8$ lattice. For the $16\times 16$ system these
parameters are 
$a=8.03 \pm 0.33$ and $b=-0.580 \pm 0.04$ respectively. It therefore seems that $\lambda$ depends on the 
system size
rather weakly.

During the W-L simulation as the modification factor becomes smaller, the changes in the density of the state 
profile also becomes finer. We denote by $\Delta g_k(E_i)$ the change in the logarithm of the density of states
resulting from the $k^{th}$ iteration and $E_i$ is the bin energy. The plots of $\Delta g_k(E_i)$ against $E_i$
for four iterations in 
the $8\times8$, $p^2=50$ model is shown in fig. 5. As expected, the fluctuation in  $\Delta g$ decreases with the 
decrease in $f$. These diagrams reveal a little more information than that is evident from the  $\Delta H_k^j$ curves in fig. 1,
 as the fluctuation spectrum is here available as a function of the energy bin number, rather than the fluctuation
summed over the bins. This is an important factor in every simulation since $\Delta H_k^j$ can saturate at a given value 
in spite of not being distributed uniformly over all bins. That this does not happen for this system,
 particularly as $f$ gets reduced,
 is apparent from these diagrams. However we have noticed that 
for bigger systems this behaviour of $\Delta g_k(E_i)$ is violated.

Before presenting results of FSS in the $p^2=50$ model, we outline the approach of Lee and Kosterlitz
\cite {lk1,lk2} who had worked out a method for doing FSS in systems with a first order phase transition.
When simulating an unknown system, two issues need to be resolved. One is the nature of phase transition and 
the other is the determination of various thermodynamic quantities. Lee and Kosterlitz proposed a 
convenient method for the determination of the order of the phase transition and this can be applied to MC 
simulations in systems having linear dimension less than the correlation length. For a temperature driven
first order transition in a finite system of volume $L^d$ with periodic boundary condition one needs to compute
a quantity $N(E;\beta,L)$ which is the histogram count of the energy distribution. The $p^2=50$ model has a 
characteristic double peak structure for $N(E;\beta,L)$ in the neighbourhood of the transition temperature. The
two peaks of $N$ at $E_1(L)$ and $E_2(L)$ corresponding respectively to the ordered and disordered phases are
separated by a minimum at $E_m(L)$. A free energy like quantity is defined as
\begin{equation}
A(E;\beta,L,\mathcal N)=-\ln N(E;\beta,L)
\end{equation}
where $\mathcal N$ is the number of configurations generated. The quantity $A(E;\beta,L,\mathcal N)$ differs
from the free energy $F(E;\beta,L)$ by a temperature and $\mathcal N$ dependent additive quantity. A bulk free
energy barrier can therefore be defined as 
\begin {equation}
\Delta F(L)=A(E_m;\beta,L,\mathcal N)-A(E_1;\beta,L,\mathcal N)
\end{equation}
It may be noted that at the transition temperature, $A(E_1;\beta,L,\mathcal N)=A(E_2;\beta,L,\mathcal N)$
and $\Delta F$ is independent of $\mathcal N$. For a continuous transition $\Delta F(L)$ is independent of
$L$ and for a temperature driven first order transition it is an increasing function of $L$, even when $L$
is smaller than the correlation length, $\xi$, prevailing at the system at the transition temperature. If one 
is in a region where $L$ is much greater than $\xi$, then $\Delta F$ obeys the scaling relation \cite{binder}
\begin{equation}
\Delta F \sim L^{d-1}
\end{equation}

For FSS analysis in the $p^2=50$ model we have used systems of 
size $L \times L$ with the linear dimension $L=16, 32, 48, 64$ and $80$. We observe that while using 
the method 'a' for the W-L 
algorithm, the termination of the simulation for a given $f$ at the pre-determined value of MC sweeps, as is 
necessary to reach the saturation value of the histogram fluctuation, does not work for $L>32$. This is because,
 a large number of bins, in the low energy range are not visited at all. The problem becomes more severe as the
system size increases. The error in the density of states resulting from the energy range not being sampled at
all leads to wrong values of thermodynamic variables. This means that, mere specification of a pre-assigned 
value of the necessary number of MC sweeps, without monitoring if each bin is adequately sampled, does not 
provide us with a method which is an alternative to the flatness test of the histograms. For clarity we present
in table 1, the energy ranges for different system sizes, which are not sampled at all for MC sweep 
ranging upto the
estimated pre-assigned number.

Next we turn to results of FSS using the methods 'b' and 'c'. Figure 6
shows the specific heat per particle plotted against the dimensionless temperature for $L$ ranging from $8$ to
$80$ and the sharpness of the $C_v$-peaks grow rapidly with increasing system size. The free energy $A$ is 
plotted against energy/particle for all lattice sizes in figure 7. The growth of a free energy barrier between 
two wells can be seen. The data for both fig. 6 and 7 have been obtained by using method method 'b'. In fig. 8
we show the finite size scaling of the peak height of $C_v$ obtained by the two methods 
where $C_v$ has been plotted against
$L^2$. In accordance with first order scaling laws \cite{binder}, 
the behaviour of $C_v$ vs $L^2$ is linear and the difference
between the results obtained from the two methods of sampling is insignificant. In figure 9, we have plotted
the free energy barrier $\Delta F$ at the transition temperatures vs the lattice size $L$. Using method 'b' we
get a good linear fit (denoted by filled circles in the figure)
which is expected from the analysis of Lee and Kosterlitz. However,
 for the method 'c' the linear behaviour of $\Delta F$ vs $L$ is not obtained (denoted by filled triangles
in the figure)
although $\Delta F$ remains a 
monotonically increasing function of $L$, which also characterizes a first order phase transition
\cite {lk1,lk2}.

In figure 10 we present the finite size scaling behaviour of the transition temperatures for different values
of $L$ obtained from i) peak position of $C_v$ and ii) fine tuning of free energy with temperature till 
two equally deep minima are obtained. We find that the fit of the transition temperature 
vs $1/L^2$ is linear, thus obeying the
finite size scaling rules for a first order transition. When we fit the data, we get two sets of straight lines
for the two methods we used. Table 2 shows the thermodynamic limits of the transition temperatures thus obtained
and the difference between the two methods appears to be insignificant.

We end this letter by noting down our conclusion. In the first part of the work we have found that, in the two 
planar spin models with continuous energy spectrum, the fluctuation in the energy histogram, after an initial
increase, saturates to a value proportional to  $1/\sqrt(\ln f)$ where $f$ is the modification factor in the
W-L sampling. Since the error in the resulting density of states depends on this saturation value, it may seem that
W-L sampling should be carried out only till the saturation value of histogram fluctuation is attained. We have
shown that this does not work even for systems of moderate size, as an energy range near the ground state is
not sampled at all. In the second part of our work, we have carried out the W-L simulation of the $p^2=50$
model for lattices of size upto $80 \times 80$ using two schemes, where the minimum number of visits to
each bin are  $1/\sqrt(\ln f)$ and  $1/\ln f$ respectively. The second scheme, although requiring a vast amount
of computation, particularly as $f$ gets smaller, does not give results which are significantly different from
the method employing $1/\sqrt(\ln f)$ visits. All the expected FSS behaviour for first order phase transition
has been seen to be obeyed in the simulations we have done and the $1/\sqrt(\ln f)$ method seems to be adequate. It
may be noted that even the $1/\sqrt(\ln f)$ visit scheme in W-L simulation in lattices of bigger size seems
to be impractical in view of the huge computer time necessary. This is a serious problem with the W-L algorithm 
when applied to continuous lattice spin models.

\noindent {\bf Acknowledgment}

	The authors acknowledge the award of a CSIR (India) research grant 03(1071)/06/EMR-II which enabled us
to acquire four IBM X 226 servers, with which the work was done. One of us (SS) acknowledges the award
of a fellowship from the same project.

\begin{figure}[tbh]
\begin{center}
\psfrag{x}{$j$}
\psfrag{y}{$\Delta H_k^j$}
\psfrag{f2}{$\ln f_k=10^{-2}$}
\psfrag{f3}{$\ln f_k=10^{-3}$}
\psfrag{f4}{$\ln f_k=10^{-4}$}
\psfrag{f5}{$\ln f_k=10^{-5}$}
\psfrag{a}{$(a)$}
\psfrag{b}{$(b)$}
\psfrag{c}{$(c)$}
\psfrag{d}{$(d)$}
\psfrag{k=5}{$k=5$}
\psfrag{k=9}{$k=9$}
\psfrag{k=13}{$k=13$}
\psfrag{k=17}{$k=17$}
\begin{tabular}{cc}
      \resizebox{80mm}{!}{\rotatebox{-90}{\includegraphics[scale=0.6]{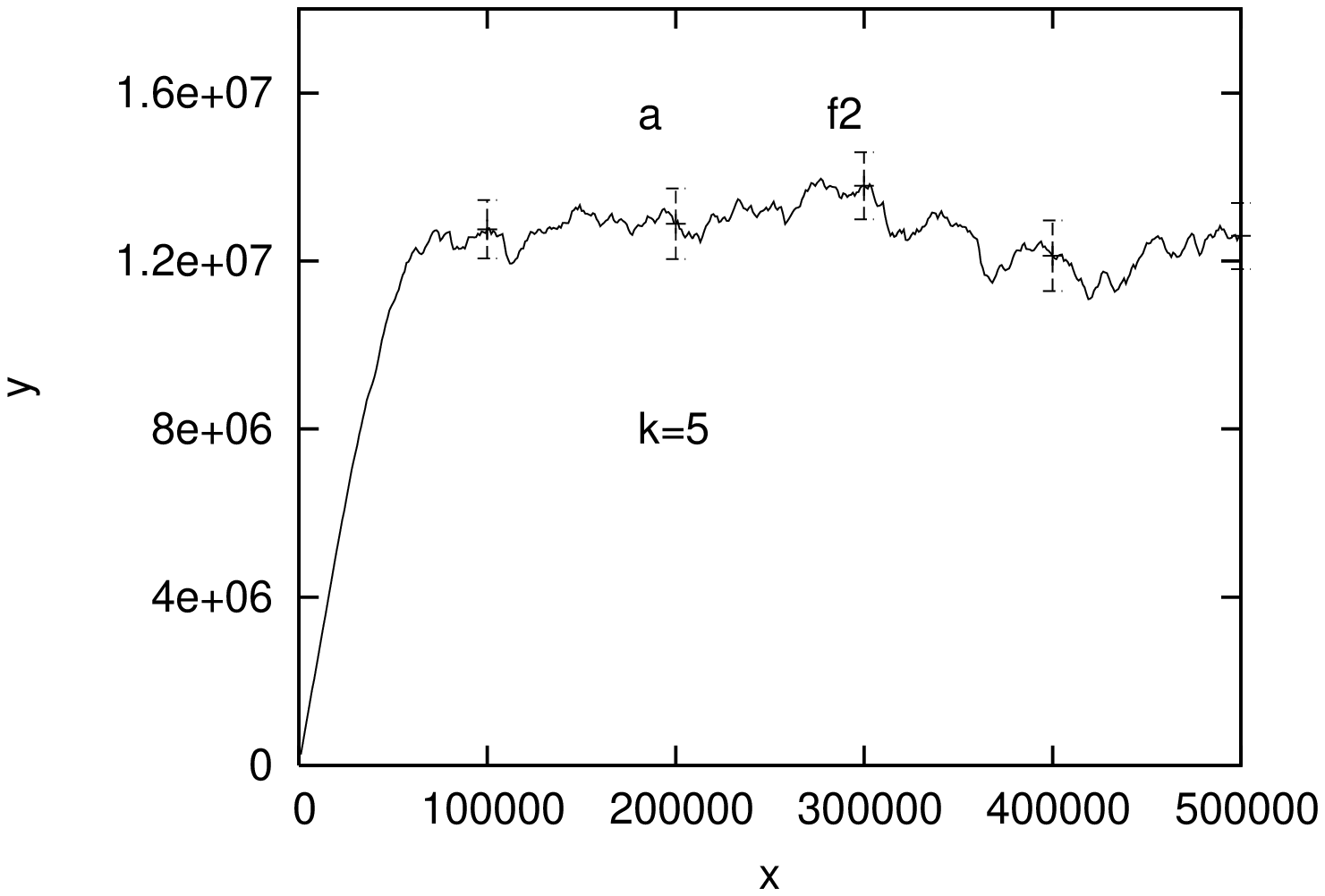}}} &
      \resizebox{80mm}{!}{\rotatebox{-90}{\includegraphics[scale=0.6]{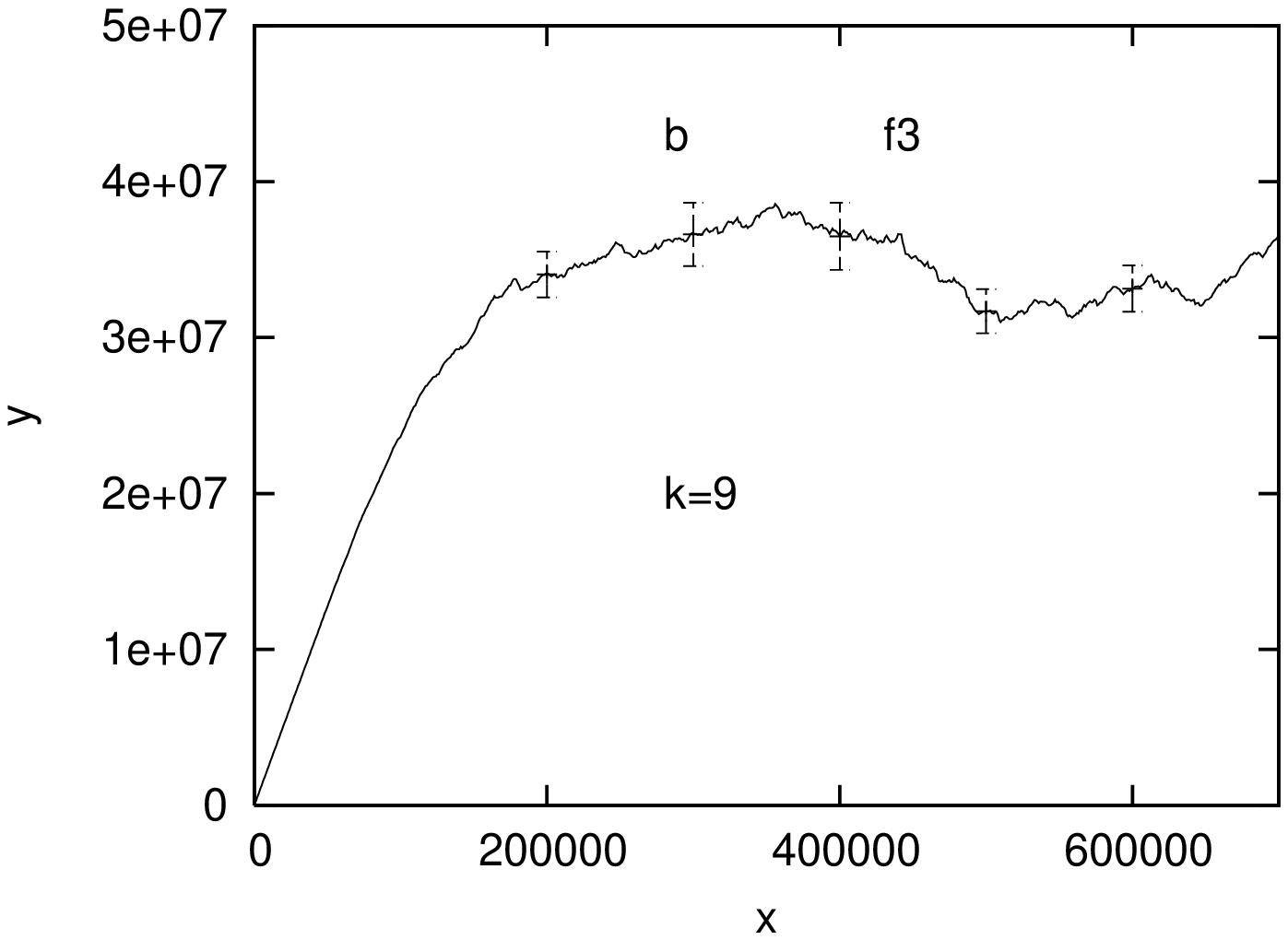}}} \\
      \resizebox{80mm}{!}{\rotatebox{-90}{\includegraphics[scale=0.6]{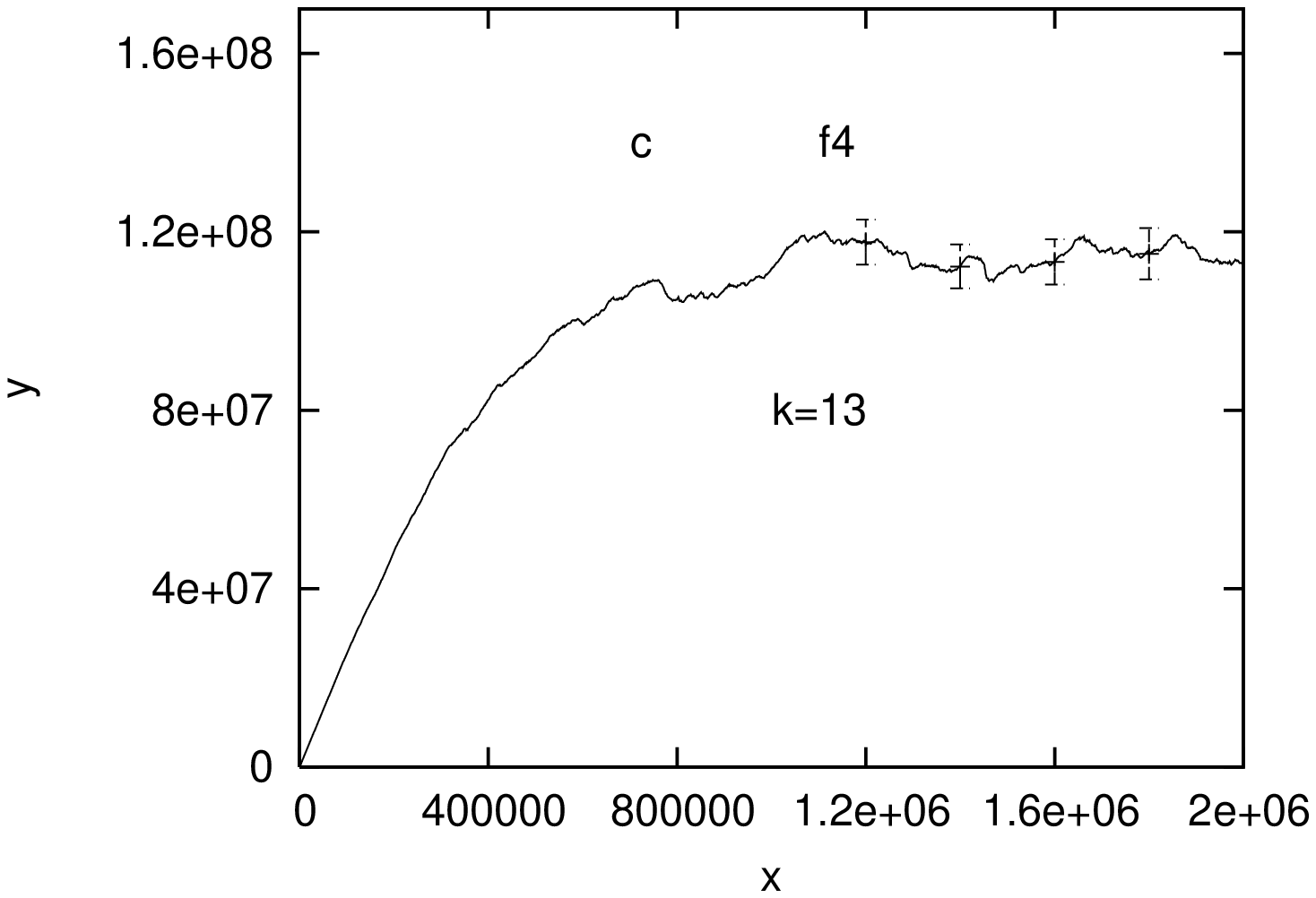}}} &
      \resizebox{80mm}{!}{\rotatebox{-90}{\includegraphics[scale=0.6]{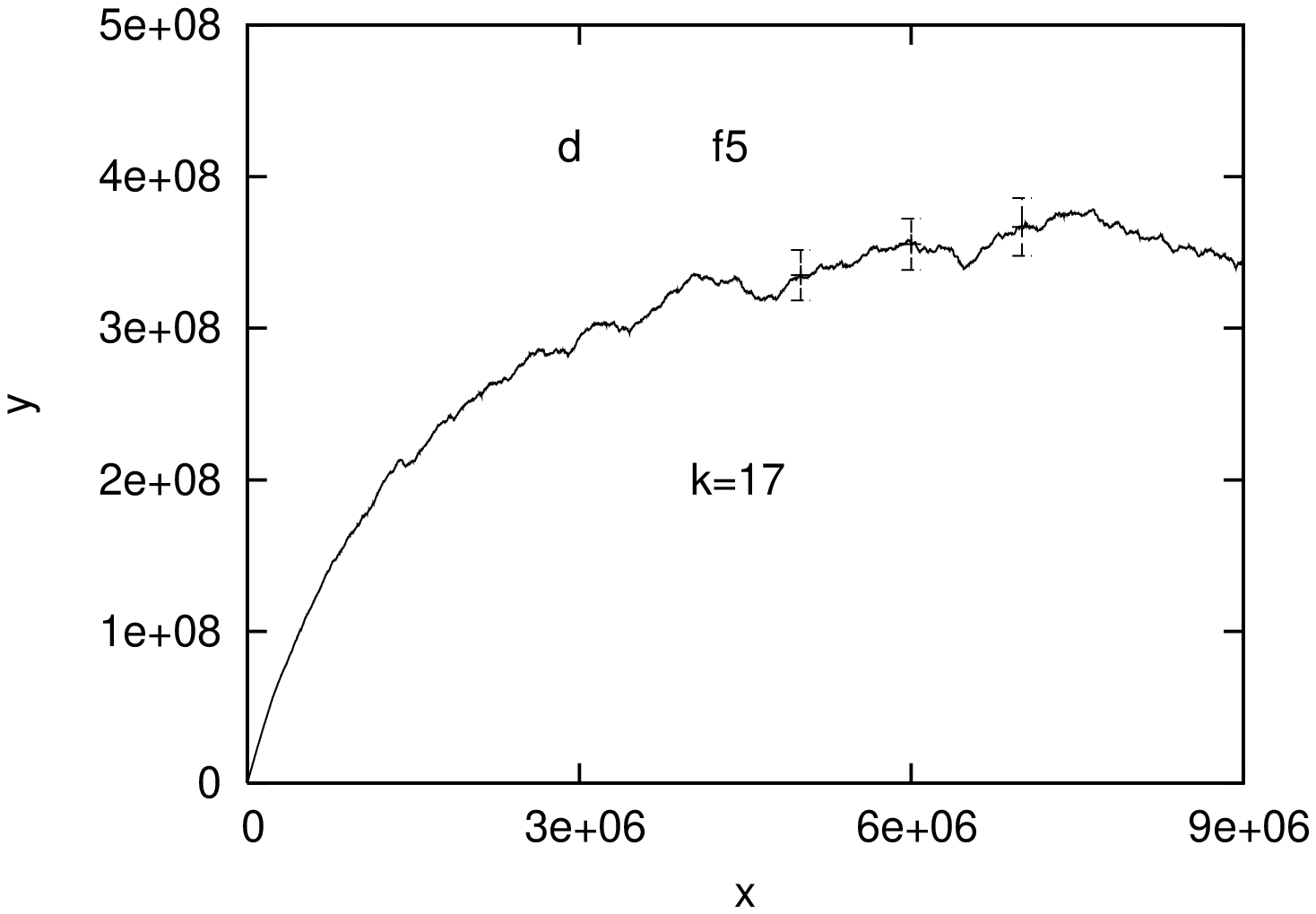}}} \\
    \end{tabular}
\end{center}
\caption{}
\label{fig1}
\end{figure}

\begin{figure}[tbh]
\begin{center}
\resizebox{100mm}{!}{\includegraphics[scale=1.2]{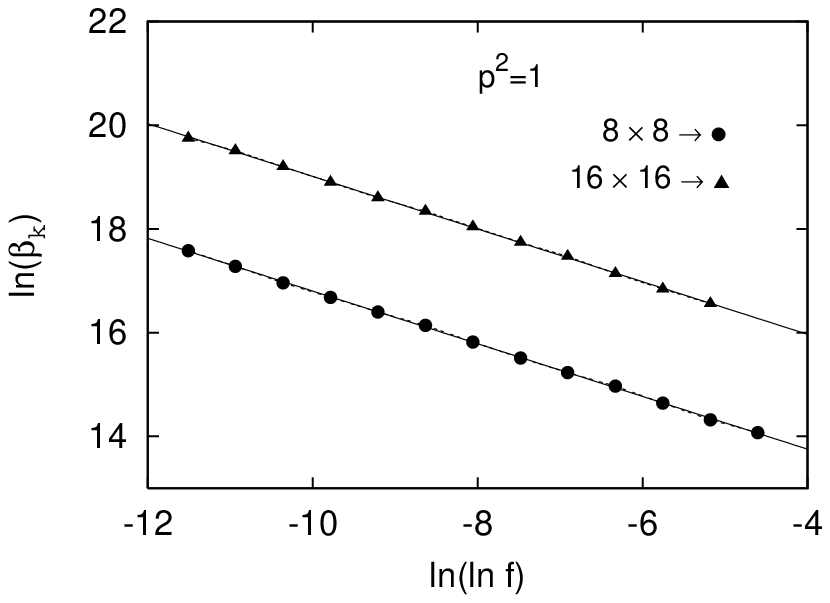}}
\end{center}
\caption{}
\label{slopep1}
\end{figure}

\begin{figure}[tbh]
\begin{center}
\resizebox{100mm}{!}{\includegraphics[scale=1.2]{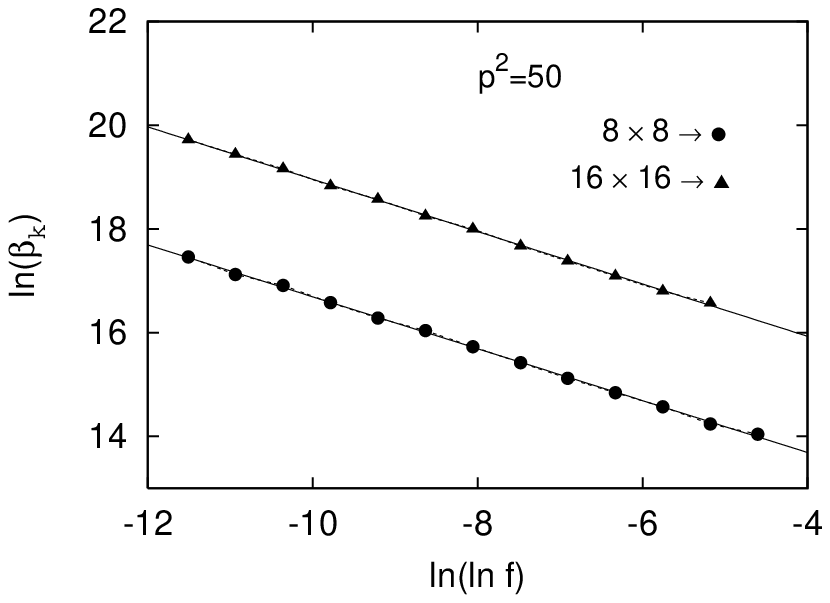}}
\end{center}
\caption{}
\label{slopep1}
\end{figure}

\begin{figure}[tbh]
\begin{center}
\psfrag{\ln(\ln f)}{$\ln(\ln f)$}
\psfrag{\ln(MCS)}{$\ln \lambda$}
\psfrag{8mul8--+}{$8 \times 8 \rightarrow +$}
\psfrag{16mul16--+}{$16 \times 16 \rightarrow \times$}
\psfrag{p}{$p^2=50$}
\resizebox{80mm}{!}{\includegraphics[scale=1.2]{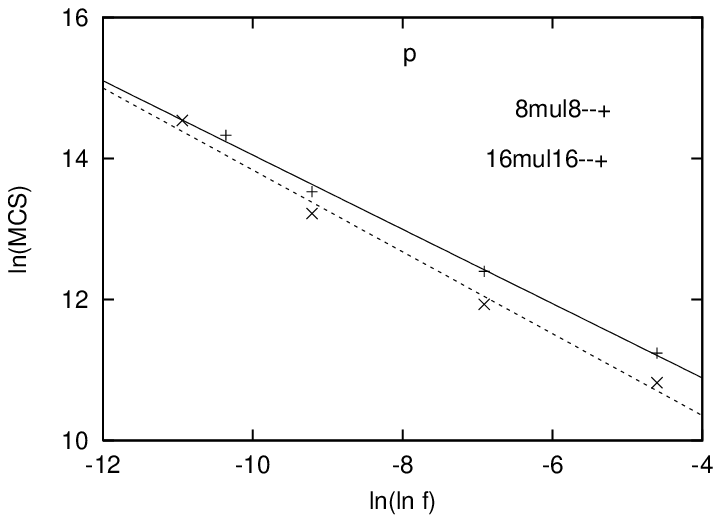}}
\end{center}
\caption{}
\label{fvsmcss}
\end{figure}

\begin{figure}[tbh]
\begin{center}
\psfrag{binindex}{$E_i$}
\psfrag{delta(lng)}{$\Delta g_k(E_i)$}
\psfrag{f6}{$\ln f_k=10^{-6}$}
\psfrag{f3}{$\ln f_k=10^{-3}$}
\psfrag{f4}{$\ln f_k=10^{-4}$}
\psfrag{f5}{$\ln f_k=10^{-5}$}
\psfrag{a}{$(a)$}
\psfrag{b}{$(b)$}
\psfrag{c}{$(c)$}
\psfrag{d}{$(d)$}
\psfrag{k=21}{$k=21$}
\psfrag{k=9}{$k=9$}
\psfrag{k=13}{$k=13$}
\psfrag{k=17}{$k=17$}
\begin{tabular}{cc}
      \resizebox{80mm}{!}{\rotatebox{-90}{\includegraphics[scale=0.6]{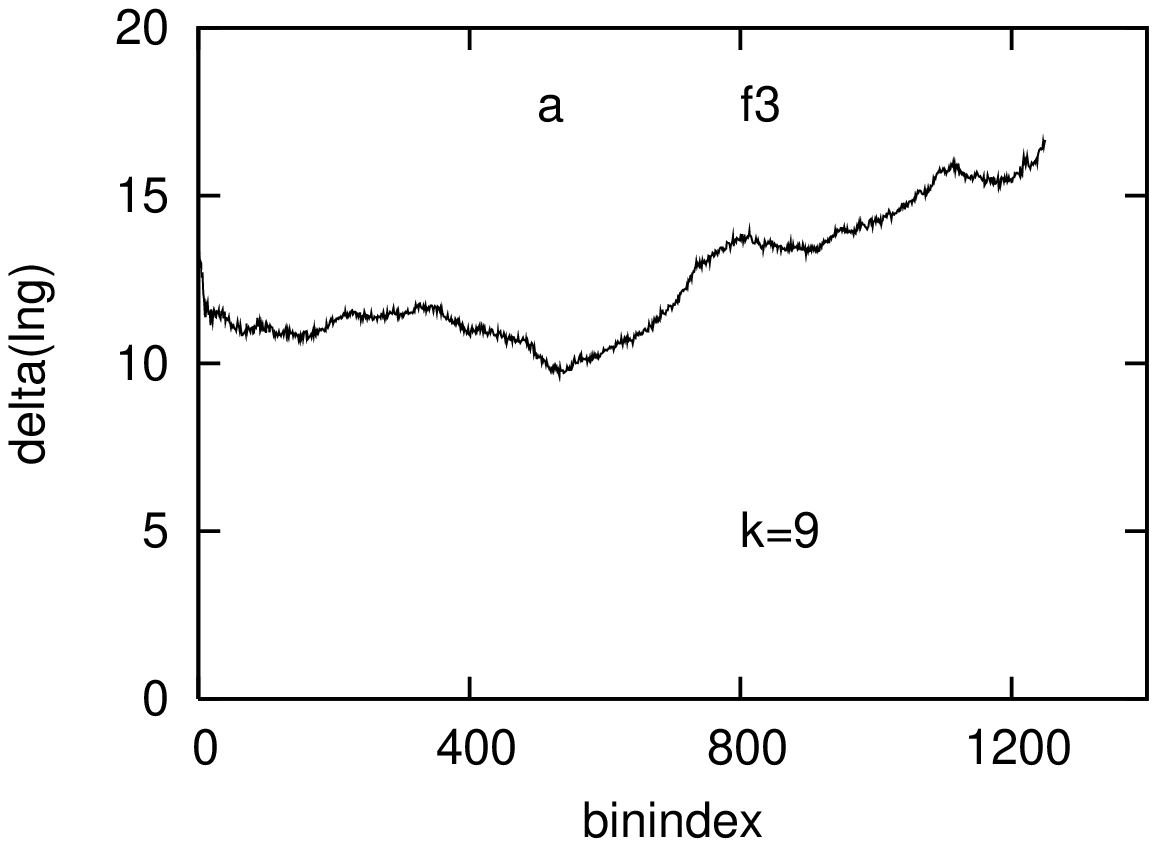}}} &
      \resizebox{80mm}{!}{\rotatebox{-90}{\includegraphics[scale=0.6]{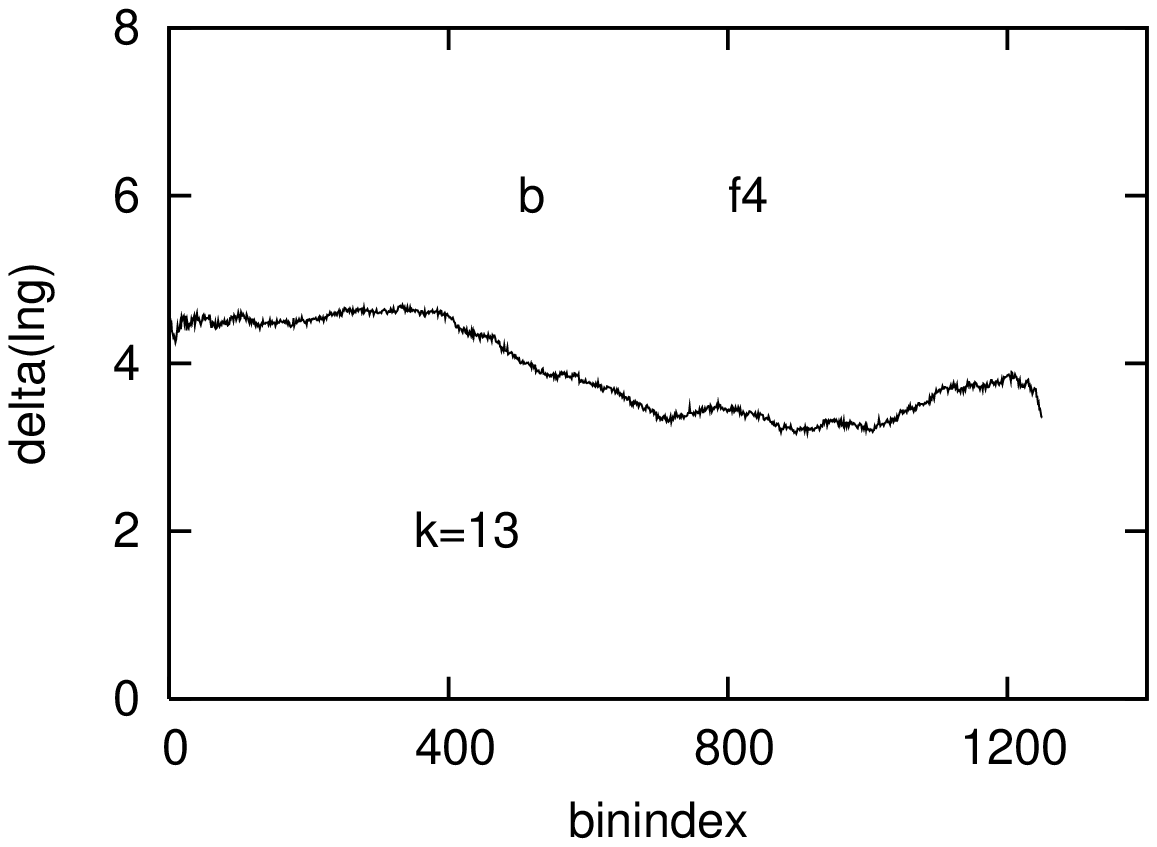}}} \\
      \resizebox{80mm}{!}{\rotatebox{-90}{\includegraphics[scale=0.6]{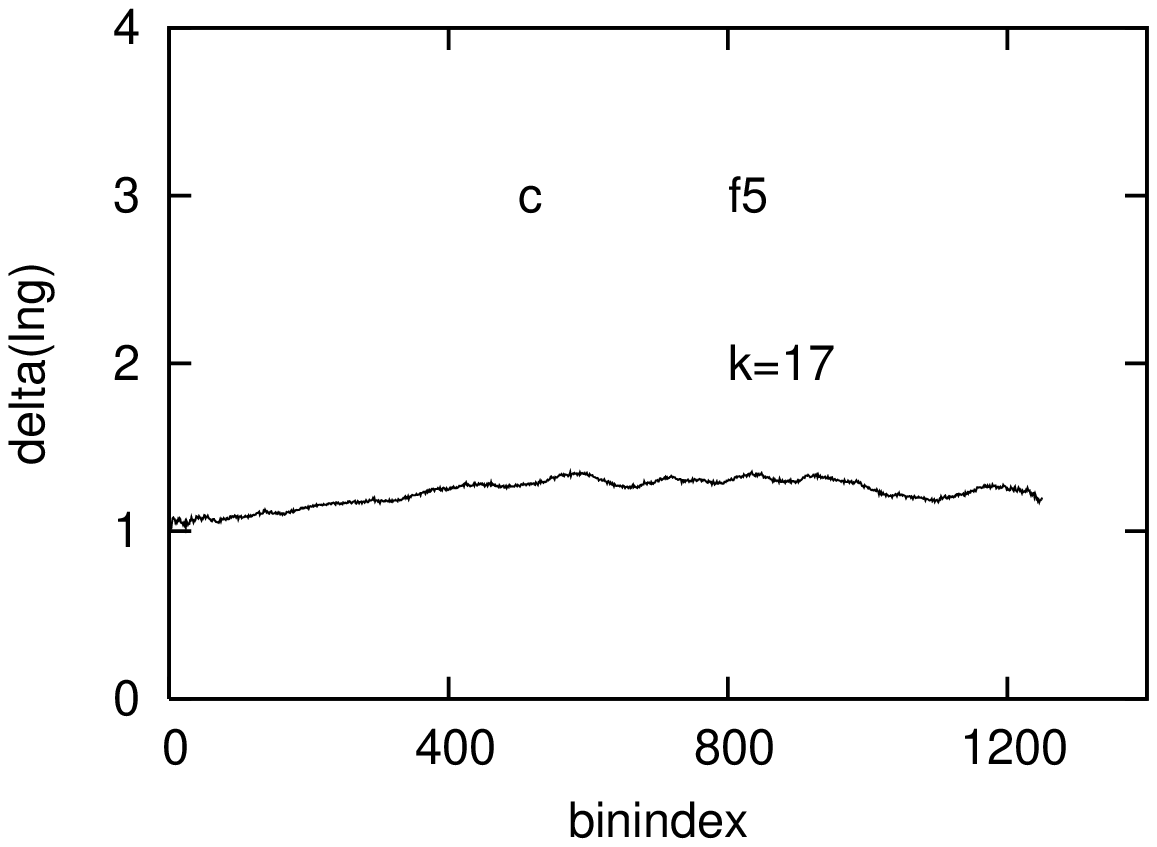}}} &
      \resizebox{80mm}{!}{\rotatebox{-90}{\includegraphics[scale=0.6]{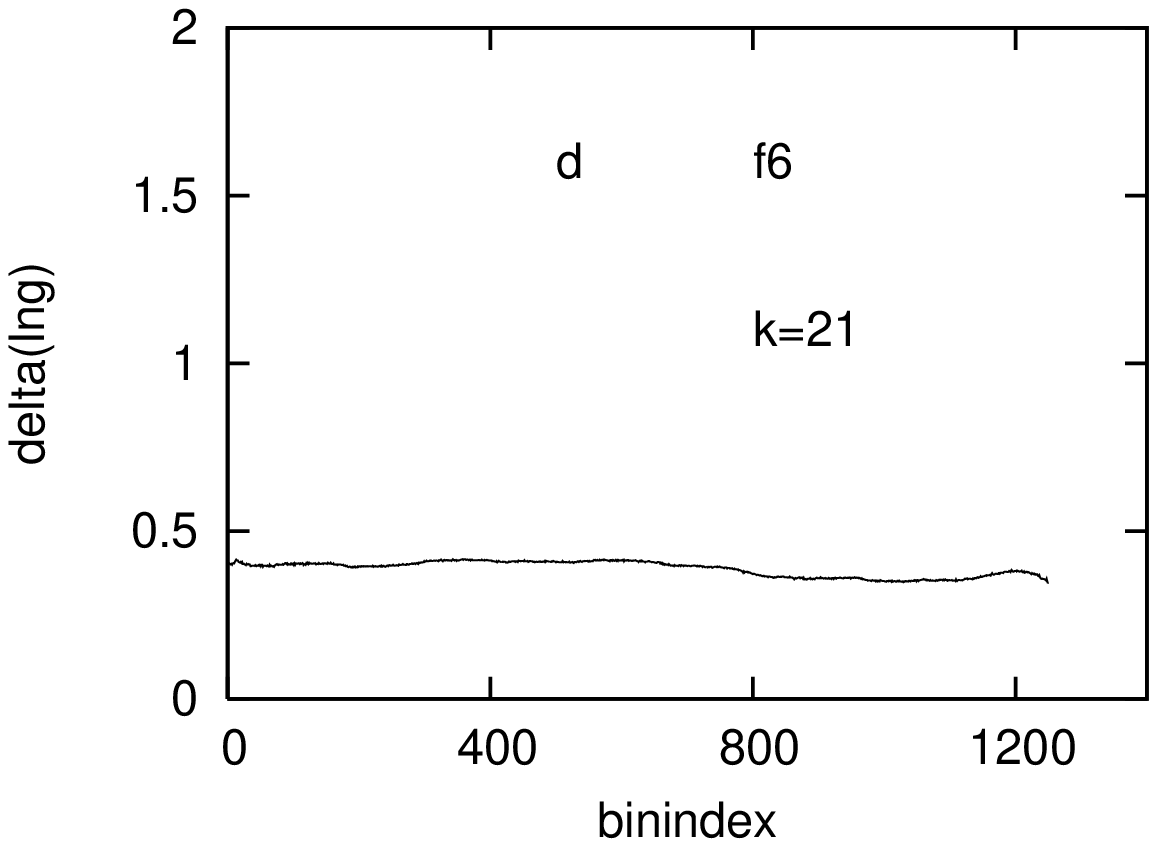}}} \\
    \end{tabular}
\end{center}
\caption{}
\label{fig5}
\end{figure}

\begin{figure}[tbh]
\begin{center}
\psfrag{Cv}{$C_v$}
\psfrag{T}{$T$}
\resizebox{80mm}{!}{\includegraphics[scale=1.2]{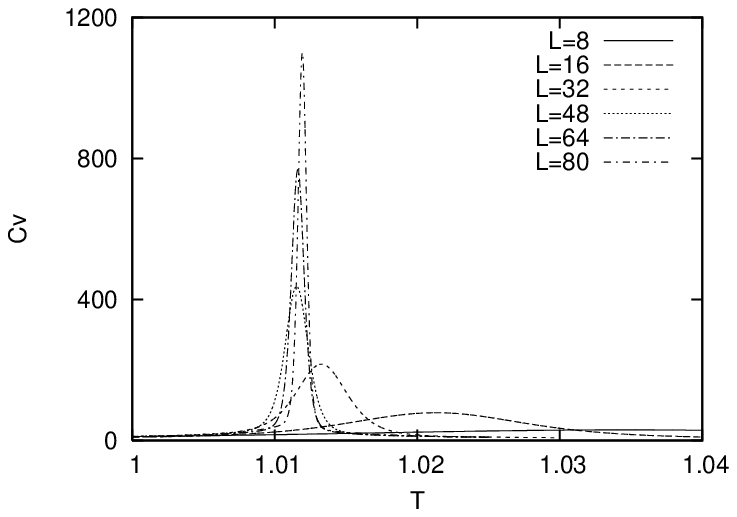}}
\end{center}
\caption{}
\label{fig6}
\end{figure}

\begin{figure}[tbh]
\begin{center}
\psfrag{freeenergy}{$A$}
\psfrag{E*}{$E/particle$}
\resizebox{80mm}{!}{\includegraphics[scale=1.2]{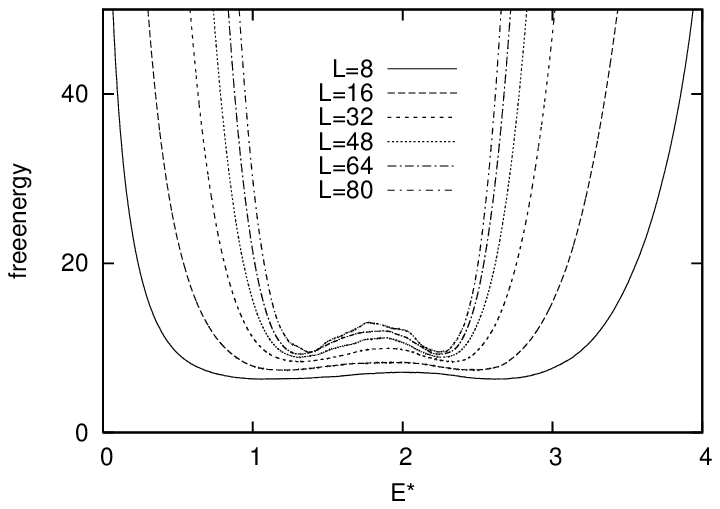}}
\end{center}
\caption {}
\label{fig7}
\end{figure}

\begin{figure}[tbh]
\begin{center}
\psfrag{Cv/particle}{$C_v/particle$}
\psfrag{L2}{$L^2$}
\resizebox{100mm}{!}{\includegraphics[scale=1.2]{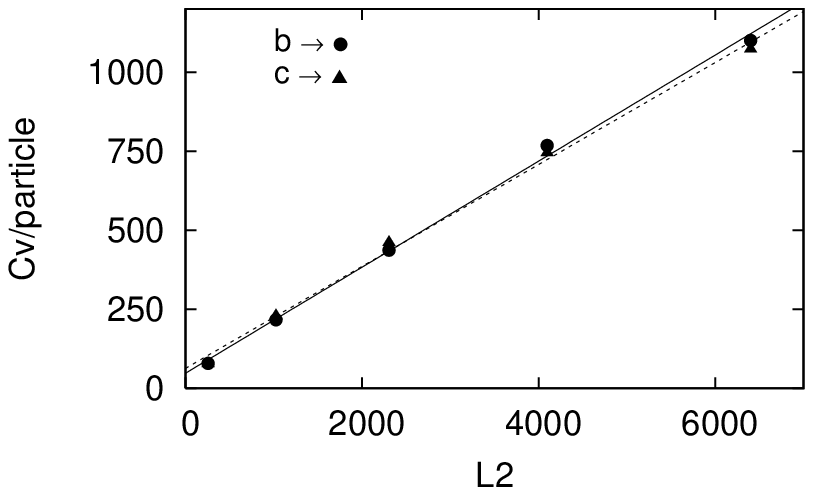}}
\end{center}
\caption{}
\label{fig8}
\end{figure}

\begin{figure}[tbh]
\begin{center}
\psfrag{\Delta F}{$\Delta F$}
\psfrag{L}{$L$}
\resizebox{100mm}{!}{\includegraphics[scale=1.2]{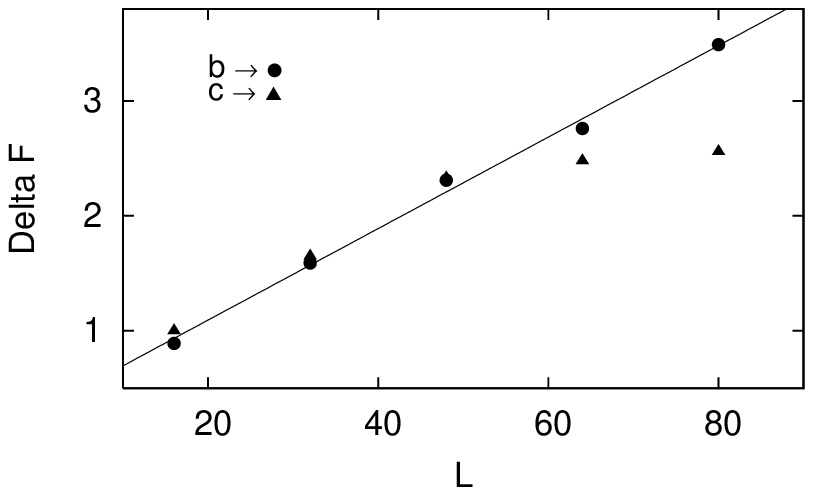}}
\end{center}
\caption{}
\label{fig9}
\end{figure}

\begin{figure}[tbh]
\begin{center}
\psfrag{Tc}{$T_c$}
\psfrag{L2}{$L^{-2}$}
\resizebox{100mm}{!}{\includegraphics[scale=1.2]{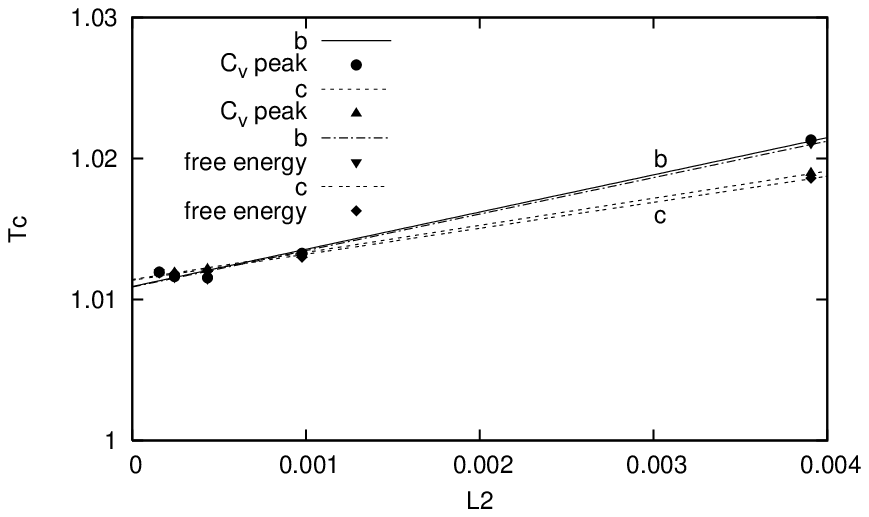}}
\end{center}
\caption{}
\label{fig10}
\end{figure}



\clearpage

\begin{table}[!ht]
\begin{center}\
\begin{tabular}{|c|c|c|}
\hline
$L$ & Energy range used & Energy range not\\
 &in simulation &visited at all\\
\hline
$8$ & $3-253$ & Entire range is sampled\\
\hline
$16$ & $10-1004$ & Entire range is sampled\\
\hline
$32$ & $10-4050$ & $10-363$ \\
\hline
$64$ & $50-16300$ & $50-8243$ \\
\hline
\end{tabular}
\end{center}
   \caption{The energy ranges used in the simulation for the $p^2=50$ model for system sizes of linear 
dimension $L$ using the method of pre-assigned MC sweeps (method 'a'). The third column shows the energy range which is
not visited by the random walk process.}
\end{table}

\begin{table}[!ht]
\begin{center}\
\begin{tabular}{|c|c|c|c|c|}
\hline
&\multicolumn{2}{c|}{method 'b'} &\multicolumn{2}{c|}{method 'c'}\\
\cline{2-5}
& $C_v$ & $\Delta F$ &$C_v$ & $\Delta F $ \\
\hline
Transition &1.01092  &1.01088 & 1.01142 & 1.01136\\ 
temperature &$\pm 0.00027$ &$\pm 0.00030$ &$\pm 0.00008$ &$\pm 0.00010$\\
for $L=\infty$ & & & &\\
\hline
\end{tabular}
\end{center}
   \caption{The thermodynamic limit of transition temperatures for the $p^2=50$ model obtained in the  
methods 'b' and 'c'. Results obtained from scaling of the specific heat peak position and fine tuning of free energy 
till the well depths are equal, are shown in the table along with the errors in the linear fits.}
\end{table}

\clearpage
\noindent {\bf Figure Captions:} \\

\noindent Figure 1: The histogram fluctuations $\Delta H_k^j$ for the $k^{th}$ iteration are plotted
against the MC sweep index $j$ for the $16\times16$ lattice and $p^2=50$. The values of $\ln f$ are indicated
in the figures. The histograms are averaged over $50$ independent simulations. The error bars are shown
in the diagram.
\vskip .1in

\noindent Figure 2: Logarithm of the saturation values of the histogram fluctuation, $\ln (\beta_k)$
is plotted against $\ln (\ln f)$ for $8\times 8$ and $16\times 16$ systems for $p^2=1$.
 The slopes of the two linear fits are given in the text. The error bars are of the dimensions smaller
than the symbols
used for plotting.
\vskip .1in

\noindent Figure 3: The plots similar to those in fig.2 for $p^2=50$. The error bars are of the dimensions 
smaller than the symbols
used for plotting.
\vskip .1in

\noindent Figure 4: Plot of $\ln \lambda$, the logarithm of the MC sweeps necessary to reach saturation of the histogram fluctuation
\vskip .1in

\noindent Figure 5: The change in the value of the logarithm of the density of states, $\Delta g_k(E_i)$
that occurs during an iteration is plotted against the bin energy $E_i$. The changes are shown for
four iterations, with the values of $\ln f$ inscribed in the diagrams. The results are for the
$8\times 8$ lattice and $p^2=50$.
\vskip .1in

\noindent Figure 6: Specific heat per particle obtained by method 'b'
plotted against the dimensionless temperature for different lattice
sizes.
\vskip .1in

\noindent Figure 7: The free energy $A$ obtained by method 'b'
plotted against bin energy per particle for all lattice sizes. For clarity 
only a restricted energy range is shown in the plot.
\vskip .1in

\noindent Figure 8: Finite size scaling of the peak height of $C_v$ per particle obtained by the methods 'b' and 'c' and plotted
against $L^2$. The error bars are of the size smaller than the symbols
used for plotting.
\vskip .1in

\noindent Figure 9: Finite size scaling of the free energy barrier at the transition temperatures by the two methods
plotted against $L$. The linear fit is for method 'b' only.
The error bars are of the size smaller than the symbols
used for plotting
\vskip .1in

\noindent Figure 10: Finite size scaling of the transition temperatures obtained from i) peak
position of $C_v$ and ii) fine tuning of free energy with temperature plotted against $L^{-2}$.
The straight lines joining data points are guide to eye. We have used results obtained from both methods.
The upper pair of the straight lines are obtained from method 'b' and the lower pair from method 'c'.
The error bars are of the dimensions smaller than the symbols
used for plotting.


\begin{thebibliography}{99}
\bibitem{wl} F. Wang and D. P. Landau, Phys. Rev. Lett.  86 (2001) 2050;
Phys. Rev. E 64 (2001) 056101.
\bibitem{va1} C. Yamaguchi and Y. Okabe, J. Phys. A 34 (2001) 8781.
\bibitem{va2} M. S. Shell, P. G. Debenedetti and A. Z. Panagiotopoulos,
J. Chem. Phys. 119 (2003) 9406.
\bibitem{va3} N. Rathore and J. J. de Pablo,
J. Chem. Phys. 124 (2002) 7225.
\bibitem{va4} Tushar S. Jainand and J. J. de Pablo,
J. Chem. Phys. 116 (2002) 7238.
\bibitem{jaya} D. Jayasri, V. S. S. Sastry and K. P. N. Murthy, Phys. Rev. E 
 72 (2005) 036702. 
\bibitem{cont} P. Poulain, F. Calvo, R. Antoine, M. Broyer and Ph. Dugourd, 
Phys. Rev. E 73 (2006) 056704.
\bibitem{local} C. Zhou, T. C. Schulthess, Stefan Torbrugge and D. P. Landau,
 Phys. Rev. Lett. 96 (2006) 120201.
\bibitem{cont1} Q. L. Yan, R. Faller, and J. J. de Pablo, J. Chem. Phys.
 116 (2002) 8745.
\bibitem{zhou} C. Zhou and R. N. Bhatt, Phys. Rev. E  72 (2005) 025701.
\bibitem{lol} H. K. Lee, Y. Okabe and D. P. Landau, Comp. Phys. Comm., 
 175 (2006) 36. 
\bibitem{im1} P. Dayal, S. Trebst, S. Wessel, D. Wurtz, M. Troyer, S. Sabhapandit 
and S. N. Coppersmith, Phys. Rev. Lett. 92 (2004) 097201.
\bibitem{im2} R. E. Belardinelli and V. D. Pereyra, Phys. Rev. E 75 (2007) 046701.
\bibitem{im3} Alexander N. Morozov and S. H. Lin, Phys. Rev. E 76 (2007) 026701.
\bibitem{ds} E. Domany, M. Schick and R. H. Swendsen, Phys. Rev. Lett. 52 (1984) 1535.
\bibitem{him} J. E. Van Himbergen, Phys. Rev. Lett. 53 (1984) 5.
\bibitem{apskr} Abhijit Pal and Soumen Kumar Roy, Phys. Rev. E 67 (2003) 011705 
\bibitem{pel} N. V. Priezjev and Robert A. Pelcovits, Phys. Rev. E 64 (2001) 031710.
\bibitem{mr1} Z. Zhang, O. G. Mouritsen and M. J. Zuckermann, Phys. Rev. Lett 69 (1992) 2803.
\bibitem{mr2} Z. Zhang, M. J. Zuckermann and O. G. Mouritsen Mol. Phys. 80 (1993) 1195.
\bibitem{mr3} C. Zannoni, Mol. Cryst. Liq. Cryst. 49 (1979) 247.
\bibitem{mr4} C. Chiccoli, P. Pasini, P. Biscarini and C. Zannoni, Mol. Phys. 65 (1988) 1505.
\bibitem{mr5} C. Chiccoli, P. Pasini, and C. Zannoni, Int. J. Mod. Phys. B 11 (1997) 1937.
\bibitem{metro} N. Metropolis, A. W. Rosenbluth, M. N. Rosenbluth, A. H. Teller and E. Teller, J. Chem. Phys.
 21 (1953) 1087.
\bibitem{ro1} H. J. Knops, Phys. Rev. Lett. 39 (1977) 766; J. V. Jose, L. P. 
Kadanoff, S. Kirkpatrick and D. R. Nelson, Phys. Rev. B 16, (1977) 1217.
\bibitem{lk1} Jooyoung Lee and J. M. Kosterlitz, Phys. Rev. B 43 (1990) 3265.
\bibitem{lk2} Jooyoung Lee and J. M. Kosterlitz, Phys. Rev. Lett 65 (1990) 137.
\bibitem{binder} Monte Carlo Methods in Statistical Physics, edited by K. Binder (Springer-Verlag, 
 1986).

\end{thebibliography}
\end{document}